\documentclass[aps,prb,twocolumn,superscriptaddress,letterpaper,longbibliography]{revtex4-1}

\usepackage{graphicx}
\usepackage{bm}
\usepackage{amssymb}
\usepackage{color}
\usepackage{amsmath}
\usepackage{ulem}

\def\GJ{\textcolor{black}}

\begin{document}

\title{Delocalization of topological edge states}
\author{Weiwei Zhu}
\thanks{These authors contribute equally.}
\affiliation{Department of Physics, National University of Singapore, Singapore 117542, Singapore}

\author{Wei Xin Teo}
\thanks{These authors contribute equally.}
\affiliation{Department of Physics, National University of Singapore, Singapore 117542, Singapore}

\author{Linhu Li}
\email{lilh56@mail.sysu.edu.cn}
\affiliation{Guangdong Provincial Key Laboratory of Quantum Metrology and Sensing $\&$ School of Physics and Astronomy, Sun Yat-Sen University (Zhuhai Campus), Zhuhai 519082, China}

\author{Jiangbin Gong}
\email{phygj@nus.edu.sg}
\affiliation{Department of Physics, National University of Singapore, Singapore 117542, Singapore}

\begin{abstract}
The non-Hermitian skin effect (NHSE) in non-Hermitian lattice systems depicts the exponential localization of eigenstates at system's boundaries.  It has led to a number of counter-intuitive phenomena and challenged our understanding of bulk-boundary correspondence in topological systems.  This work aims to investigate how the NHSE localization and topological localization of in-gap edge states compete with each other,
with several representative static and periodically driven 1D models, whose topological properties are protected by different symmetries.
The emerging insight is that at critical system parameters, even topologically protected edge states can be perfectly delocalized.  In particular, it is discovered that this intriguing delocalization occurs if the real spectrum of the system's edge states falls on the same system's complex spectral loop obtained under the periodic boundary condition.  We have also performed sample numerical simulation to show that such delocalized topological edge states can be safely reconstructed from time-evolving states.  \GJ{Possible applications of delocalized topological edge states are also briefly discussed}.

\end{abstract}

\maketitle
\section{Introduction}
The so-called non-Hermitian skin effect (NHSE) has been recognized as one of the seminal findings in non-Hermitian lattice systems~\cite{yao2018,yao2018a,MartinezAlvarez2018,kunst2018,lee2019,yokomizo2019,lee2019a,longhi2019,herviou2019,zhang2020,li2020a,li2020,chunhui2020helical,weidemann2020,scheibner2020,hofmann2019,brandenbourger2019,fei2019damping}.
Due to NHSE,  eigenstates of a non-Hermitian system tend to exponentially pile up at its boundaries.
This challenges our conventional thinking of bulk-boundary correspondence (BBC)
\cite{gong2018,takata2018,FoaTorres2019,ghatak2019,brzezicki2019,deng2019,edvardsson2019,ezawa2019,kunst2019,wang2019,fei2019realspace,Longhi2020,yi2020,kawabata2020,yang2020}, which often connects eigenstate localization with band topology alone.   For example, \GJ{ normally the generalized Brillouin zone approach 
\cite{yao2018,yokomizo2019,lee2019a}}
has to be introduced to remove the complications of NHSE so that we can correctly describe the topological properties of non-Hermitian systems by non-Bloch topological invariants
\cite{kawabata2020,yang2020,li2019geometric,CH2020,imura2019,jin2019,ghatak2019a,helbig2020,xiao2020}.
On the other hand,
NHSE itself has a topological origin arising from point-gapped spectrum with spectral winding on the complex plane~\cite{kawabata2019,okuma2020,zhang2020a}. With NHSE demonstrated in several experiments~\cite{weidemann2020,scheibner2020,hofmann2019,brandenbourger2019,ghatak2019a,helbig2020,xiao2020}, it is of even more interest to explore new faces of NHSE in topological non-Hermitian systems.




In this work, our focus is to investigate the localization/delocalization of topological edge states in several representative non-Hermitian systems.
Intuitively, the behavior of topological edge states depends on both topological localization and NHSE as two competing factors.   Topological localization can be largely connected with the width of a topological band gap, whereas NHSE is of a different origin. How these two factors compete then presents an interesting subject to study.  The main job of this work is to identify details of such a competition.  This can become highly nontrivial in systems with more than one characteristic length scales of NHSE, e.g., in a periodic driven system where the NHSE can be challenging to analyze \cite{zhang2020} or in a three-band system with a complicated spectrum on the complex plane.   As a key finding, we are able to identify a key condition to observe completely delocalized topological edge states.  For one model, we have also performed explicit simulations to show that the profile of such delocalized topological edge states can be reconstructed from time-evolving states.

\GJ{The main message of this work arises from our computational observations rather than a formal theory.  What we shall uncover here is one overlooked and yet very useful aspect of the spectrum of non-Hermitian systems under the periodic boundary condition (PBC).   Indeed, in previous studies, the PBC spectrum of a non-Hermitian system is largely used, via the generalized Brillouin zone approach 
\cite{yao2018,lee2019a,yokomizo2019,kawabata2020,yang2020,li2019geometric,CH2020}, 
to predict and understand the {\it bulk} spectrum of the system under the open boundary condition (OBC).  In this type of routine by which one unravels the non-Hermitian pumping \cite{CH2020}, the properties of the possible topological edge states are not in the picture.   It is hence remarkable that this work shows that the PBC spectrum is actually essential to understand under what conditions the topological edge states may become completely delocalized.}

This work is organized as follows.  Section II starts our analysis with a non-Hermitian \GJ{Su-Schrieffer-Heeger (SSH)}  model, in connection with the emergence of exceptional points and half-integer Bloch topological invariants.  This is followed by Sec.III, where a periodically driven (Floquet) SSH model is studied.  Such a Floquet two-band model is richer, because it has now two quasi-energy band gaps.   Results there enhance our analysis in the static case, insofar as the delocalization of topological edge states associated with both gaps are observed and it occurs at the exceptional point of the two-band system.  {Besides,  some numerical simulations presented there also show that the profile of such delocalized edge states can be reconstructed from 
time-evolving states.}
{With these results, one might infer that the delocalization of topological edge states is a property of exceptional band topology.} However, in Sec.~IV we are able to identify a more general condition using a three-band system without the kind of chiral symmetry in the previous two examples.
The magic condition for delocalization of topological edge states turns out to be the same in the all examples studied, namely, it occurs when the spectrum of the system under the periodic boundary condition touches that of the topological edge states.  Sec.V concludes this paper, \GJ{with a brief discussion on possible applications of delocalized edge states}.


\section{Delocalized topological edge state in non-Hermitian SSH model}

\begin{figure}
\includegraphics[width=\linewidth]{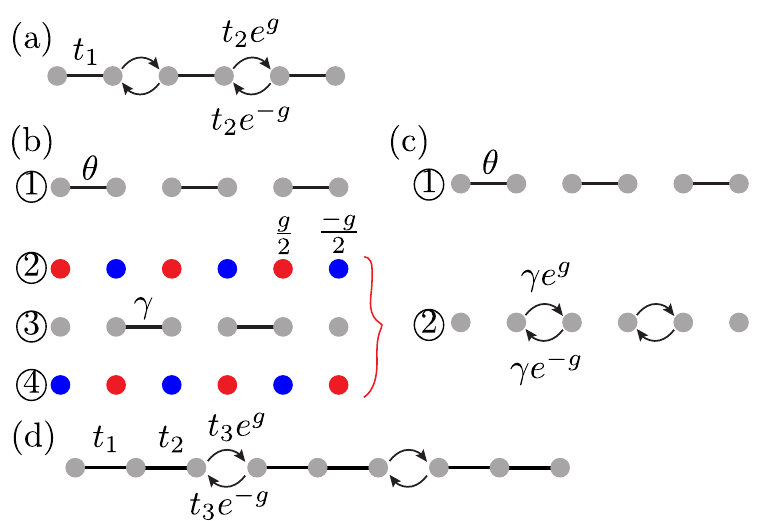}
\caption{(a) Non-Hermitian SSH model. $t_1$ is the intracellular couplings and asymmetric couplings are introduced into the intercellular couplings described by $t_2$ and $g$. (b)(c) Non-Hermitian Floquet SSH model. (b) The non-Hermitian Floquet SSH model is composed of four steps. In odd steps, the lattice is fully dimerized with the site only coupled with one of its nearest sites. In even steps, the lattice has alternative gain and loss. (c) The equivalent two-step model, with asymmetric couplings in even steps. The intracellular couplings is $\theta$ and the intercellular couplings is $\gamma$. $g$ is the gain strength. (d) Non-Hermitian three-band extended SSH model. $t_{1,2}$ are the intracellular couplings and asymmetric couplings are introduced to the intercellular couplings described by $t_3$ and $g$. This model is conjugated-inversion symmetric when $t_1=t_2$. }
\label{staticmodel}
\end{figure}

We first study the non-Hermitian SSH model shown in Fig.~\ref{staticmodel}(a) which is composed of two sublattices. The intracellular couplings are $t_1$ and the intercellular couplings are asymmetric, i.e. $t_2e^{g}$ for right couplings and $t_2e^{-g}$ for left couplings. The Hamiltonian in momentum space is

\begin{equation}\label{eq1}
  H(k)=\left(\begin{array}{cc}
               0 & t_1+t_2e^{i(k-ig)} \\
               t_1+t_2e^{-i(k-ig)} & 0
             \end{array}
  \right),
\end{equation}
satisfying a chiral symmetry $\sigma_zH(k)\sigma_z=-H(k)$ that protects a $Z$-type topology in 1D.
In complex plane, $e^{ik}$ gives the Brillouin zone (BZ) as a unit circle centered at origin, and $e^{i\tilde{k}}$ with $\tilde{k}=k+ig$ gives the generalized Brillouin zone (GBZ) as another circle with the same center but a different radius of $e^{-g}$.
According to the non-Bloch bulk-boundary correspondence, the topological properties under open boundary conditions (OBCs) are described by $H(\tilde{k})$ with $\tilde{k}=k+ig$ corresponding to the GBZ, meaning that they are determined only by $t_1$ and $t_2$, but not by the value of $g$.
Namely, the system is topologically trivial when $t_1>t_2$, and nontrivial with topological edge states under OBCs when $t_1<t_2$.
{The absence and presence of these edge states are characterized by a winding number ($0$ and $1$ respectively) defined in the GBZ, which can be obtained from $2\pi i\tilde{\nu}=\oint_{\tilde{k}}d\tilde{z}/\tilde{z}$, with $\tilde{z}=t_1+t_2e^{i\tilde{k}}$.}

\begin{figure}
\includegraphics[width=\linewidth]{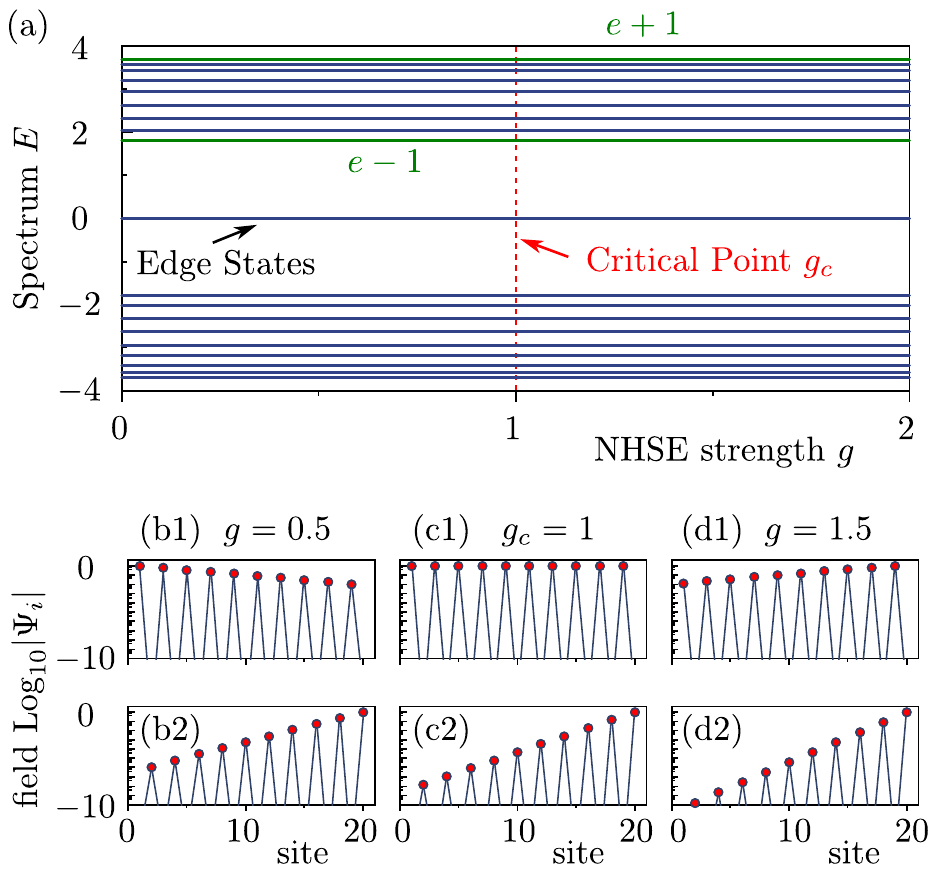}
\caption{NHSE induced delocalized topological edge state in non-Hermitian SSH model. (a) Spectrum in open boundary condition, $e-1$ and $e+1$ indicated on the panel are boundary values of the bulk spectrum.  (b1)-(d1) \GJ{Eigenstate profile} of topological edge state most occupying $A$ sublattice. (c1) The delocalized topological edge state. (b2)-(d2) \GJ{Eigenstate profile} of  topological edge state most occupying $B$ sublattice. $t_1$ and $t_2$ are set as $1$ and $e$, respectively. (b) for $g=0.5$, (c) for $g=1$ and (d) for $g=1.5$.}
\label{staticSpectrum}
\end{figure}

We show an example in Fig.~\ref{staticSpectrum}(a) with $t_1=1$ and $t_2=e$. Without loss of generality, we only consider $t_1$, $t_2$ and $g$ with positive values. The nonzero $g$ introduces NHSE into the system and all bulk states are localized at right edge.

\begin{figure}
\includegraphics[width=\linewidth]{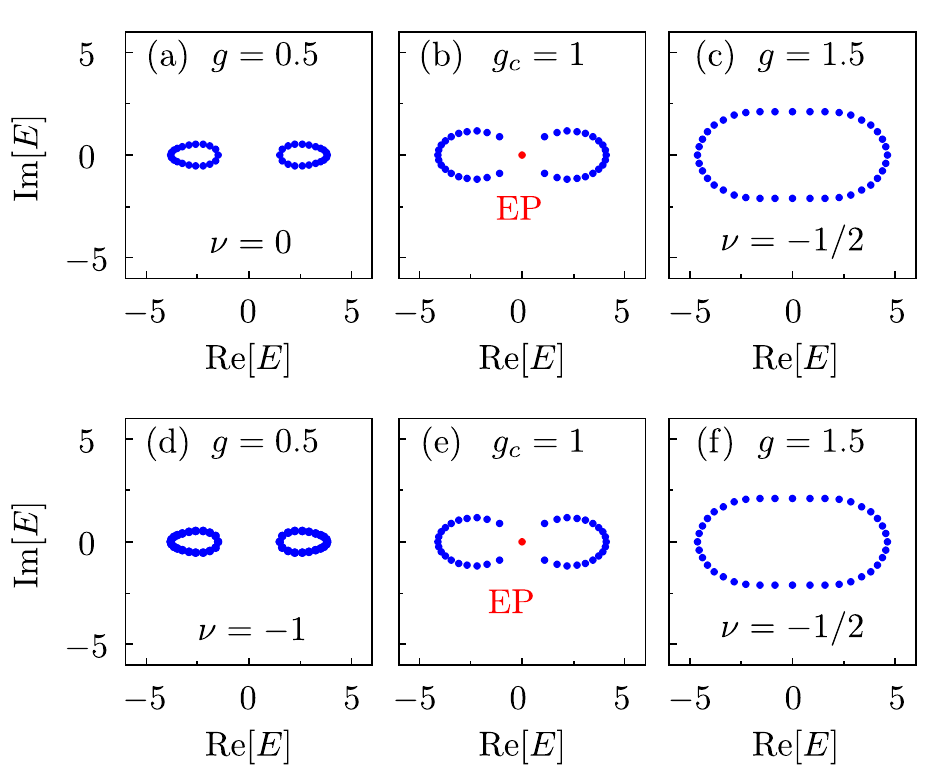}
\caption{Spectrum of non-Hermitian SSH model in periodic boundary condition and EPs. (a)-(c) for topological trivial state with $t_1=e$ and $t_2=1$. (d)-(f) for topological nontrivial state with $t_1=1$ and $t_2=e$. For both cases the critical point is at $g_c\equiv|\ln(t_1/t_2)|=1$. (c)(f) are gapless states with a half-integer topological invariant $\nu=-1/2$.}
\label{staticPBC}
\end{figure}

Due to the divergence of BZ and GBZ, another Bloch topological invariant can also be defined in the BZ, which can be determined by the average of two winding numbers~\cite{yin2018}
\begin{equation}\label{eq1}
  \nu=\frac{\nu_++\nu_-}{2}
\end{equation}
where $2\pi i\nu_\pm=\oint_kdz_\pm/z_\pm$, $z_\pm=t_1+t_2e^{\pm g}e^{ik}$.
Unlike the GBZ winding number which is always quantized, $\nu$ can be half-quantized when $\nu_+\neq\nu_-$. Specifically in our model, we have $\nu=-1/2$ when $g>g_c\equiv|\ln(t_1/t_2)|$.
Such a half integer means that the two bands connect to each other head-to-tail in the BZ, forming one big loop encircling an EP~\cite{Dembowski2001,gao2015,Shen2018,li2019geometric}. Fig.~\ref{staticPBC}(a)-(c) show the results for topological trivial states  with $t_1=e$ and $t_2=1$. And Fig.~\ref{staticPBC}(d)-(f) show the results for topological nontrivial states  with $t_1=1$ and $t_2=e$. Their critical points are all $g_c=1$. We notice the spectrum is gapped in Fig.~\ref{staticPBC}(a)(d) with $g<1$ and gapless in Fig.~\ref{staticPBC}(c)(f) with $g>1$. At the critical point $g_c$, the system supports EP shown in Fig.~\ref{staticPBC}(b)(e).

As the winding number $\nu$ depends on the value of $g$,
it is not expected to be directly related to the existence of topological edge states under OBCs. Indeed, both the cases with and without such edge states can accompany a half-integer $\nu=-1/2$, as shown in Fig.~\ref{staticPBC}(c)(f). On the other hand,
We notice that $g$ describes the localization strength of the NHSE, and $g_c\equiv|\ln(t_1/t_2)|$ corresponds to the topological localization strength of the edge states.
Thus the half-integer topological invariant is actually the results of strong NHSE whose strength is larger than the topological localization from band gap.
Therefore we can expect the change of localizing direction of the edge states with $g$ tuned across $g_c$, provided the two types of localizations are toward opposite directions. In other words, a transition of localizing direction of topological edge states shall occur when $\nu$ changes between an integer and a half-integer.

To give a concrete example, we illustrate the distribution of the topological edge states in Fig.~\ref{staticSpectrum}(b) to (d).
For $g<g_c$, the two edge states localized at different ends of the 1D chain, as shown in Figs.~\ref{staticSpectrum}(b1) and (b2). Note that they exhibit different localization strengths as the NHSE contributes to the same localizing direction for all eigenstates. With increasing $g$, the $A$-sublattice edge state is delocalized at $g=g_c$ [Fig.~\ref{staticSpectrum}(c1)] and localized at the other end when $g>g_c$ [Fig.~\ref{staticSpectrum}(d1)]. The localization strength of $B$-sublattice edge state is also getting stronger as shown in Figs.~\ref{staticSpectrum}(c2) and (d2).
The localization of two edge states can be described by $-g_c+g$ and $g_c+g$, negative value means localized at left edge.
Such a transition of localizing direction of topological edge states also coincides with the geometric transition of the PBC spectrum, where the two bands touch at the origin and give rise to an EP [Figs.~\ref{staticPBC}(b) and (e)].


\begin{figure*}
\includegraphics[width=\linewidth]{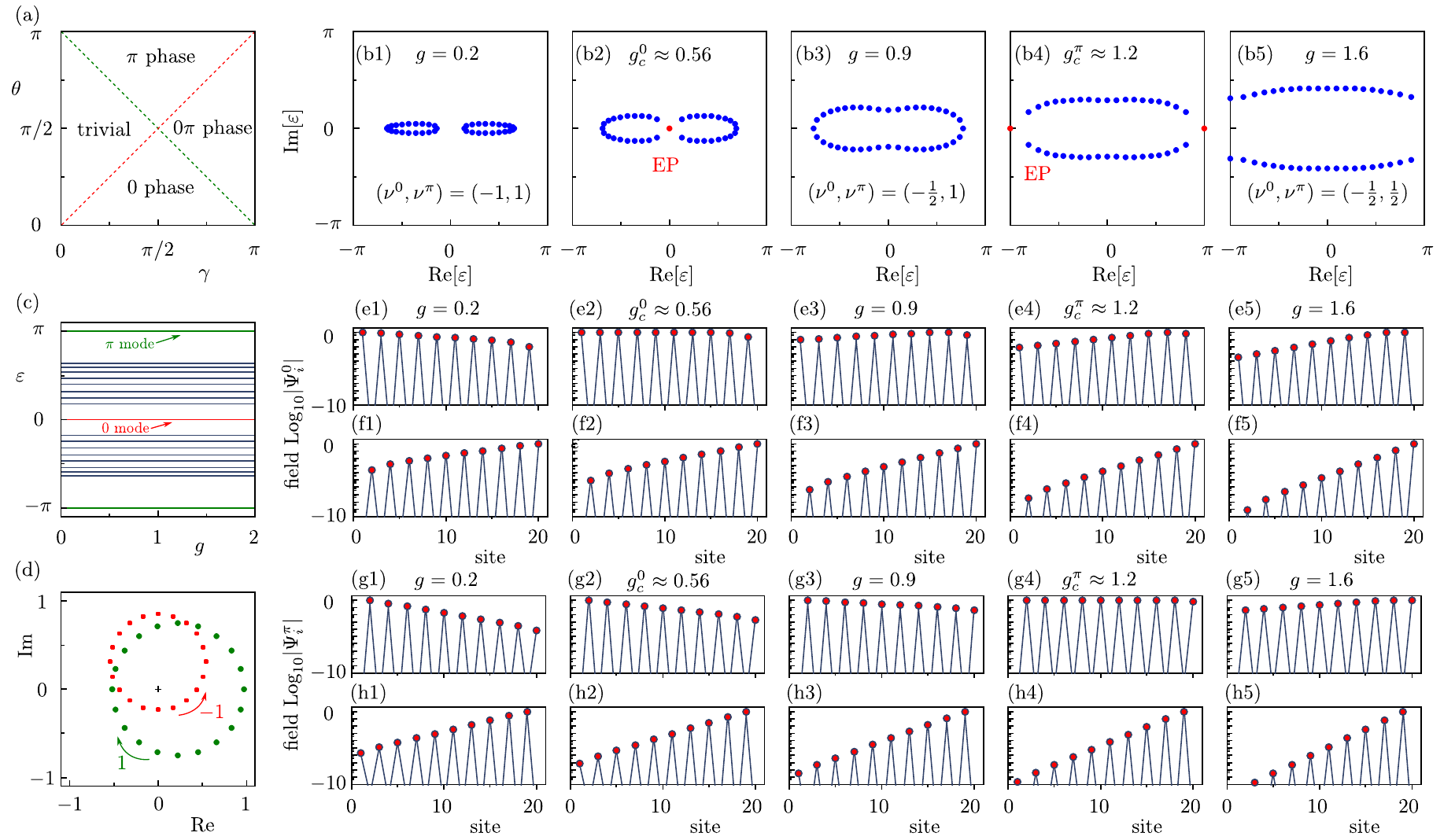}
\caption{NHSE induced delocalized topological edge state in non-Hermitian Floquet SSH model and EPs. (a) Phase diagram. There are four phases:  trivial phase, $0$ phase which supports $0$ mode, $\pi$ phase which supports $\pi$ mode and $0\pi$ phase which supports both $0$ mode and $\pi$ mode. (b) Spectrum in periodic boundary condition with different values of $g$. (c) Spectrum in open boundary condition as function of $g$. (d) The non-Bloch winding number for $0$ gap and $\pi$ gap. (e)(f) Two edge states with quasienergy $0$. (g)(h) Two edge states with quasienergy $\pi$. (e2)(g4) are delocalized topological edge states. In (b)-(h), we choose $\theta=0.6\pi$ and $\gamma=0.75\pi$.}
\label{floquetSpectrum}
\end{figure*}

\section{Delocalized topological edge state in non-Hermitian Floquet SSH model}
The non-Hermitian SSH model in Fig.~\ref{staticmodel}(a) requires asymmetric couplings which is hard to realize in experiment. Here we propose a non-Hermitian Floquet SSH model shown in Fig.~\ref{staticmodel}(b) which are composed of four steps. In odd steps, the lattice is fully dimerized with site only coupled with one of its nearest site. In even steps, the lattice has alternative gain and loss on potential. Such four-steps model is equivalent to a two-step  model with asymmetric couplings shown in Fig.~\ref{staticmodel}(c). The time dependent Bloch Hamiltonian is

\begin{eqnarray}
 H(k,t)=\left\{
\begin{array}{c}
H_a(k),\,\,0<t\leq T/2   \\
H_b(k),\,\,T/2<t\leq T
\end{array}
\right.
\label{eq2}
\end{eqnarray}

where $H_a(k)=2\theta\sigma_x$ and

\begin{equation}\label{eq3}
  H_b(k)=2\gamma\left(\begin{array}{cc}
                 0 &e^{i(k-ig)} \\
                  e^{-i(k-ig)} & 0
               \end{array}
  \right)
\end{equation}
The Floquet theory then leads us to the Floquet states $U_{t_0}(k,T)|\Psi\rangle=e^{i\varepsilon T}|\Psi\rangle$, with $U_{t_0}(k,T)\equiv \mathcal{T}\mathrm{exp}\big[i\int_{t_0}^{t_0+T} H(\tau)\,d\tau\big]$ being time evolution operator in one period, $\mathcal{T}$ is the time-ordering operator and $t_0$ is a reference time. In our case, we set the period $T=1$.

Similar to the non-Hermitian SSH model, the generalized Brillouin zone of such system is a circle with center at origin and radius $e^{-g}$. Its topological properties are determined by $\theta$ and $\gamma$, and independent with $g$. Topological phase transition can happen at $0$ or $\pi$ quasi-energy gap and splits the parameter space into four phases shown in Fig.~\ref{floquetSpectrum}(a). Those phases can be described by two non-Bloch winding numbers corresponding to $0$ gap and $\pi$ gap~\cite{Asboth2013,zhou2018}, { which are obtained from $2\pi i\tilde{\nu}^{\varepsilon}=\oint_{\tilde{k}}d\tilde{z}^{\varepsilon}/\tilde{z}^{\varepsilon}$, where $\varepsilon=\{0,\pi\}$, $\tilde{z}^{0}=i\cos\frac{\gamma}{2}\sin\frac{\theta}{2}+i\sin\frac{\gamma}{2}\cos\frac{\theta}{2}e^{i\tilde{k}}$ and $\tilde{z}^{\pi}=\cos\frac{\gamma}{2}\cos\frac{\theta}{2}-\sin\frac{\gamma}{2}\sin\frac{\theta}{2}e^{-i\tilde{k}}$.}

Similar to the static SSH model, while the non-Bloch winding number only takes integer, the Bloch winding number can take a half integer. The half-integer topological invariant comes from the competition between NHSE strength and band gap width. Differently, the Floquet model has two gaps at $0$ and $\pi$. So one needs two Bloch topological invariants to describe the system. They are determined by

\begin{eqnarray}
  \nu^{0} &=& \frac{\nu_+^{0}+\nu_-^{0}}{2} \\
  \nu^{\pi} &=& \frac{\nu_+^{\pi}+\nu_-^{\pi}}{2}
\end{eqnarray}
where $2\pi i\nu_\pm^{\varepsilon}=\oint_k dz_\pm^{\varepsilon}/z_\pm^{\varepsilon}$, $\varepsilon=0,\pi$
\begin{eqnarray}
 z_\pm^{0} &=& i\cos\frac{\gamma}{2}\sin\frac{\theta}{2}+i\sin\frac{\gamma}{2}\cos\frac{\theta}{2}e^{\pm g}e^{ik} \\
  z_\pm^{\pi} &=& \cos\frac{\gamma}{2}\cos\frac{\theta}{2}-\sin\frac{\gamma}{2}\sin\frac{\theta}{2}e^{\mp g}e^{-ik}
\end{eqnarray}

There are two critical points obtained from $|z_+^{0}||z_-^{0}|=0$ and $|z_+^{\pi}||z_-^{\pi}|=0$,
\begin{eqnarray}
  g_c^{0} &\equiv& \left|\ln\left(\tan\frac{\theta}{2}\cot\frac{\gamma}{2}\right)\right| \\
 g_c^{\pi}&\equiv& \left|\ln\left(\tan\frac{\theta}{2}\tan\frac{\gamma}{2}\right)\right|,
\end{eqnarray}
corresponding to the EPs at $0$ and $\pi$ quasienergies, and determining the transition of localizing direction of topological zero modes and $\pi$ modes respectively.
For trivial phase and $\pi$ phase, $\nu^{0}=0$ with $g<g_c^0$ and $\nu^{0}=-1/2$ with $g>g_c^{0}$. For $0$ phase and $0\pi$ phase, $\nu^{0}=-1$ with $g<g_c^0$ and $\nu^{0}=-1/2$ with $g>g_c^{0}$. For trivial phase and $0$ phase, $\nu^{\pi}=0$ with $g<g_c^\pi$ and $\nu^{0}=1/2$ with $g>g_c^{\pi}$. For $\pi$ phase and $0\pi$ phase, $\nu^{\pi}=1$ with $g<g_c^\pi$ and $\nu^{\pi}=1/2$ with $g>g_c^{\pi}$.

Next we focus on special case with $\theta=0.6\pi$ and $\gamma=0.75\pi$, which belongs to the $0\pi$ phase. The two critical points approximately equal to $g_c^0\approx0.56$ and $g_c^\pi\approx1.2$. At those critical points, the spectrum in periodic boundary condition support EP at $0$ gap and $\pi$ gap as shown in Figs.~\ref{floquetSpectrum}(b2) and (b4). Within $g<g_c^{0}$, the spectrum forms two separated loops in Fig.~\ref{floquetSpectrum}(b1). Within $g_c^0<g<g_c^\pi$, $\nu^0$ is half-integer and the spectrum is gapless at quasienergy $0$ shown in Fig.~\ref{floquetSpectrum}(b3). Within $g>g_c^\pi$, both $\nu^0$ and $\nu^\pi$ are half-integer, and the spectrum is gapless at quasienergy $0$ and $\pi$ in Fig.~\ref{floquetSpectrum}(b5).

Under OBC, the spectrum is independent of $g$, as shown in Fig.~\ref{floquetSpectrum}(c). The system supports $0$ mode and $\pi$ mode, \GJ{of which the presence can be digested through} the nonzero non-Bloch winding number (winding number defined in generalized Brillouin zone). The generalized Brillouin zone is used to remove the NHSE. Indeed, one can obtain the non-Bloch winding number by setting $g=0$ and calculating the Bloch winding number $\nu^{0}$ and $\nu^{\pi}$.
Fig.~5(d) show the nonzero non-Bloch winding numbers for $0$ gap and $\pi$ gap.

The localization of topological edge state is determined by band gap width and NHSE. The profile of two $0$ modes are shown in Figs.~\ref{floquetSpectrum}(e)(f). With $g<g_c^0$, two $0$ modes are localized at opposite edges in Figs.~\ref{floquetSpectrum}(e1)(f1). With $g>g_c^0$, two $0$ modes are localized at same sides in Figs.~\ref{floquetSpectrum}(e3)-(e5),(f3)-(f5). At critical point $g_c^0$, one $0$ mode is delocalized in Fig.~\ref{floquetSpectrum}(e2). The profile of two $\pi$ modes are shown in Figs.~\ref{floquetSpectrum}(g)(h). With $g<g_c^\pi$, two $\pi$ modes are localized at opposite edges in Figs.~\ref{floquetSpectrum}(g1)-(g3),(h1)-(h3). With $g>g_c^\pi$, two $\pi$ modes are localized at same sides in Figs.~\ref{floquetSpectrum}(g5)(h5). At critical point $g_c^\pi$, one $\pi$ mode is delocalized in Fig.~\ref{floquetSpectrum}(g4).

\begin{figure}
\includegraphics[width=\linewidth]{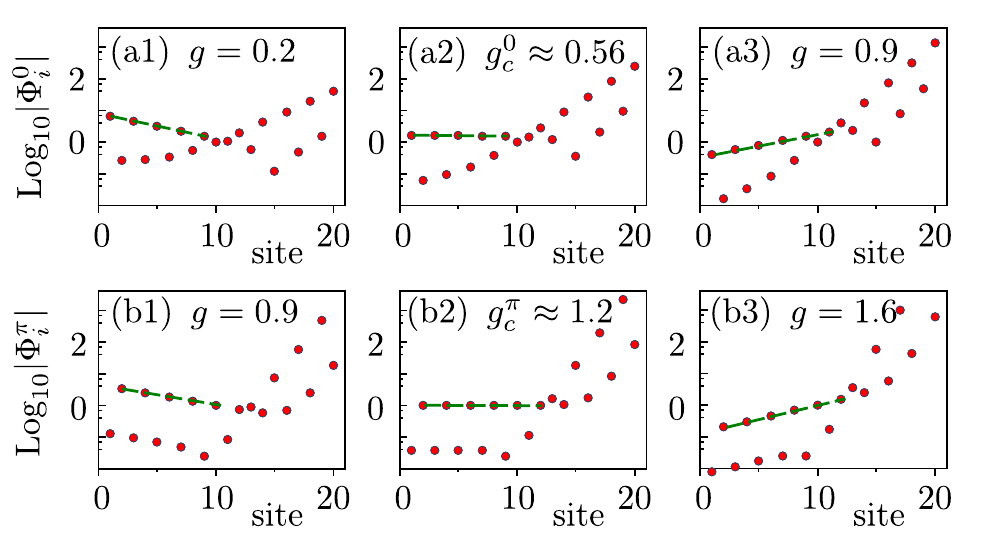}
\caption{Reconstruction of the profile of delocalized topological edge states by a simple time averaging approach.  (a) Extracted topological edge states with quasienergy $0$. (b) Extracted topological edge states with quasienergy $\pi$. $\theta=0.6\pi$ and $\gamma=0.75\pi$. The slope of dashed lines connecting the respective dots in each panel indicate the localization exponents of one of the two topological edge states, with the zero slop corresponding to delocalization of the concerned edge state.  }
\label{detecting}
\end{figure}


To end this section, we proceed to show that the delocalized topological edge states identified above can be directly relevant in experiments.  For example,  in Ref.~\cite{xiao2020} it was shown that the topological edge state in a non-Hermitian Floquet system can be extracted from the time-evolving state profile $|\Phi(t)\rangle$ by the following simple operation,

\begin{equation}
  |\Phi^{\varepsilon}(t)\rangle=\sum_{t'=0}^{t}\frac{e^{-i\varepsilon t'}}{t+1}|\Phi(t')\rangle\,\,\,(\varepsilon=0,\pi).
\end{equation}
\GJ{The underlying insight motivating this approach is the following: For a sufficiently long time $t$ used for the averaging above}, the contribution to the above expression from bulk state cancels out and only topological edge state with the specified quasienergy $\varepsilon$ retains.   This being the case, one can next extract the profile of a topological edge state by \GJ{assuming that the initial state has a finite overlap with the corresponding topological edge state.  We perform precisely such type of simulation, with the profile of the extracted topological edge states at quasienergy $0$ shown in Fig.~\ref{detecting}(a).  It is seen that} both edge states (left and right) are excited in Fig.~\ref{detecting}(a1) with $g=0.2$. At $g_c^0\approx0.56$, the left part is delocalized in Fig.~\ref{detecting}(a2). With $g=0.9>g_c^0$, the left part is also localized to right in Fig.~\ref{detecting}(a3). The extracted edge states with quasienergy $\pi$ is shown in parallel in Fig.~\ref{detecting}(b). The left part is delocalized at $g_c^\pi\approx1.2$ in Fig.~\ref{detecting}(b2).

\section{A three-band extended SSH model}
In previous sections, we have studied both static and Floquet SSH model, where a chiral symmetry ensures the absence of the third Pauli matrices, and thus allows the topological winding numbers to be well-defined for the systems.
Nonetheless,
there are different symmetries which can protect the 1D topology, such as the inversion symmetry in a generalized multi-band system \cite{hughes2010inversion,guo2015kaleidoscope,yang2015characterization,zhu2018}.
Therefore, to further explore the emergence of the delocalization of the topological edge states, we consider a three-band extended SSH model as shown in Fig.~\ref{staticmodel}(d). It is described by the following Hamiltonian:
\begin{equation}
\label{eqn:3band}
H(k) =
\begin{pmatrix}
0 & t_1 & t_3 e^{i(k-ig)} \\
t_1 & 0 & t_2 \\
t_3 e^{-i(k+ig)} & t_2 & 0
\end{pmatrix},
\end{equation}
{where $t_{1,2}$ are the intracellular couplings, and the intercellular couplings are asymmetric, with $t_3e^g$ for right couplings and $t_3e^{-g}$ for left couplings.}

In the Hermitian case with $g=0$, the three-band extended SSH model is inversion symmetric when $t_1=t_2$.
The non-Hermitian coupling strength $g$ has to break the inversion symmetry as it takes different signs toward different spacial direction. Nevertheless,
the overall Hamiltonian still satisfies a conjugated-inversion symmetry, defined as $IH(k)I^{-1}=H^{\dagger}(-k)$ with the symmetry operator $I_{ij}=\delta_{i,4-j}$.
This symmetry also coincides with the inversion symmetry when $g=0$, where the Hamiltonian becomes Hermitian, $H=H^\dagger$.
Later we shall see that the degeneracy of a pair of in-gap edge states is lifted when the symmetry is broken, suggesting a corresponding symmetry protection of the topological properties in this model.

\begin{figure}
\includegraphics[width=0.9\linewidth]{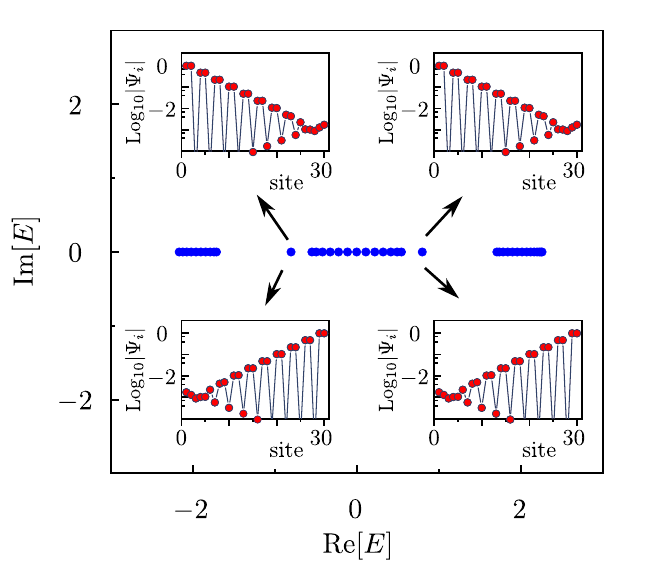}
\caption{Energy spectra of a three-band SSH system without NHSE. The blue dots are the energy spectrum under open boundary condition. The eigenstate profile of two pairs of degenerate topological edge states are shown in the insets. Parameters: $t_1=0.8$, $t_2=0.8$, $t_3=1.7$ and $g=0$.}
\label{fig:3band}
\end{figure}

We first consider the conjugated-inversion symmetric case where $t_1=t_2$.
\GJ{Analogous to} the SSH model, the topological properties in this model is also only determined by the intercellular and intracellular couplings.
{That is, the system is topologically nontrivial if $t_3>t_{1,2}$, and trivial otherwise.}
In Fig.~\ref{fig:3band}, we illustrate an Hermitian example with $t_1=t_2=0.8$, $t_3=1.7$ and $g=0$, where 2 pairs of degenerate topological edge states emerge in this system, each lies in a different band gap. The two topological edge states in each pair also localize at different ends respectively.


\begin{figure}
\includegraphics[width=\linewidth]{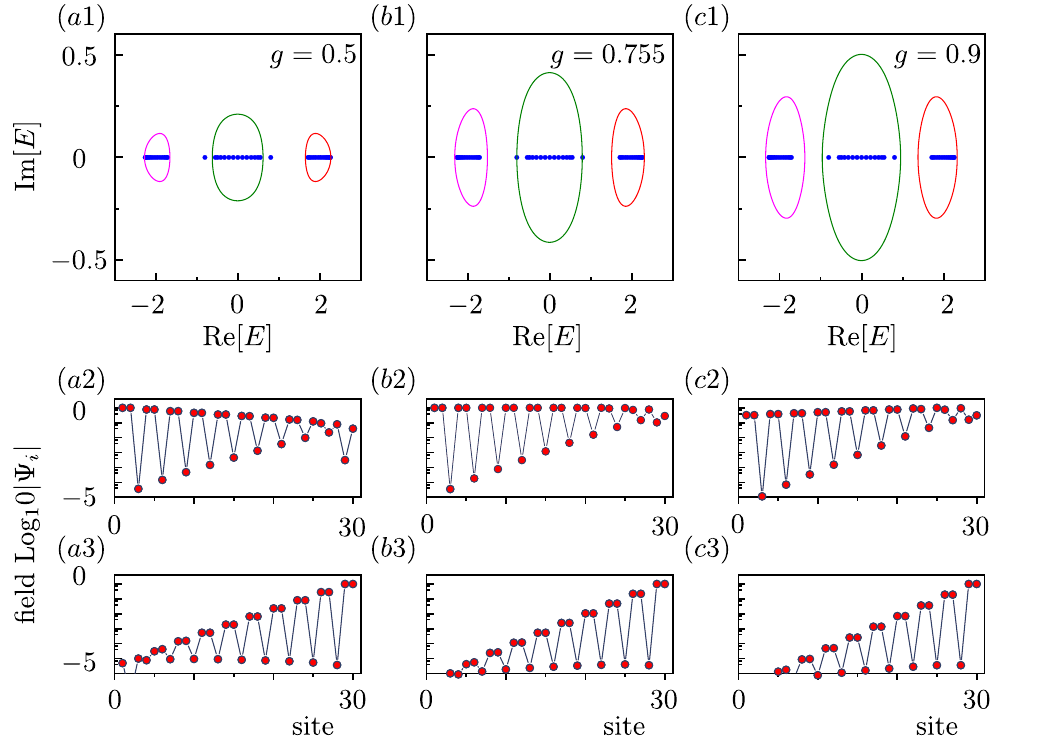}
\caption{Conjugated-inversion symmetry case. (a1)-(c1) Energy spectra in both periodic boundary conditions (pink, green, red lines) and open boundary conditions (blue dots). The loop formed by the PBC spectrum is enlarging when $g$ increases. (b1) The PBC spectrum touches the topological edge states. (a2)-(c2) Eigenstate profile of the topological edge state which originally localized at left. (b2) The delocalized topological edge state. (a3)-(c3) Eigenstate profile of the right topological edge state which originally localized at right. The negative gradient is slightly increasing when $g$ increases. Parameters: $t_1=0.8$, $t_2=0.8$, $t_3=1.7$. (a) $g=0.5$ (b) $g=0.755$ and (c) $g=0.9$. We only consider the topological edge state with positive eigen-energies.}
\label{fig:inversion}
\end{figure}

By turning on the non-reciprocal coupling $g$, the PBC spectrum will form loop and the NHSE occurs.
Increasing $g$ will enlarge the loop formed by the PBC spectrum as shown in Fig.~\ref{fig:inversion}.
The loop will eventually touch the topological edge states, which is the critical point where the topological edge states delocalized, as shown in Fig.~\ref{fig:inversion}(b2).
In contrast to the results in previous sections, exceptional point does not present in this case.

With these results obtained, it finally becomes clear that the delocalization of topological edge states does not necessarily occur at the exceptional point. As such,
the reason why previously there is a connection between delocalization and EP lies in the presence of the chiral symmetry. That is,
in the static and Floquet SSH models, the chiral symmetry ensures that the topological edge states are located at the center of band gap, which is also where the PBC spectral loop touches the real axis at the EP point. 
Therefore, in previous cases, the delocalization occurs when the topological edge states touches the EP, and hence touches the PBC spectrum.
This hence implies the following:  the general magic condition for the delocalization of the topological edge states is that the PBC spectrum touches the topological edge states.

\begin{figure}
\includegraphics[width=\linewidth]{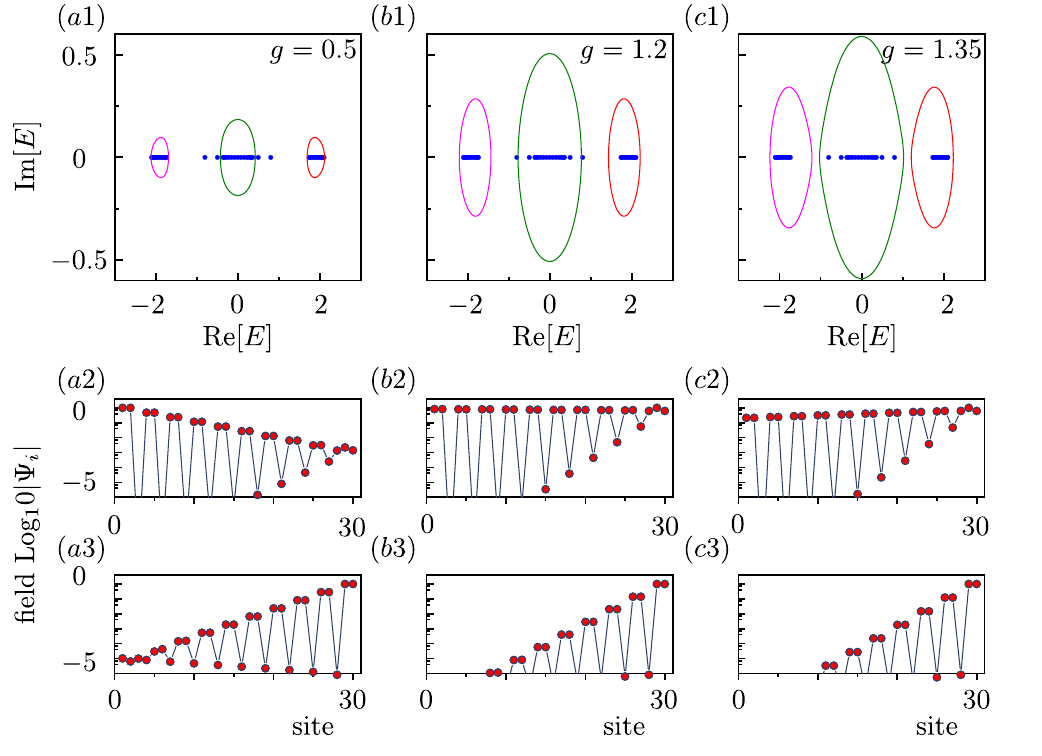}
\caption{Conjugated-inversion symmetry broken case. (a1)-(c1) Energy spectra in both periodic boundary conditions (pink, green, red lines) and open boundary conditions (blue dots). The loop formed by the PBC spectrum is enlarging when $g$ increases. (b1) The PBC spectrum touches the in-gap edge states. (a2)-(c2) Eigenstate profile of the in-gap edge state which originally localized at left with eigen energy around $E\approx0.8$. (b2) The delocalized in-gap edge state. (a3)-(c3) Eigenstate profile of the right in-gap edge state which originally localized at right with eigen energy around $E\approx0.5$. The negative gradient is slightly increasing when $g$ increases. Parameters: $t_1=0.8$, $t_2=0.5$, $t_3=1.7$. (a) $g=0.5$ (b) $g=1.2$ and (c) $g=1.35$.}
\label{fig:inversionbroken}
\end{figure}

Interestingly, this condition identified above applies not only to the topological edge states, but also non-topological in-gap edge states that are not protected by any symmetry.
For instance, if we choose $t_1\neq t_2$, the conjugated-inversion symmetry is broken.
A pair of in-gap edge states is observed in each band gap, but is no longer degenerate, as shown in Fig.~\ref{fig:inversionbroken}(a1)-(c1).
Thus they can now move and merge into different bands without closing the band gap and are no longer topologically protected.
On the other hand, the loop formed by the PBC spectrum also enlarges with the increasing $g$, and the critical point where PBC spectrum touches the edge state occurs at $g=1.2$ for the parameters we choose.
At the critical point, the edge state which is originally localized left will be delocalized, as shown in Fig.~\ref{fig:inversionbroken}(b2).

\GJ{The above seen critical point represents a critical NHSE strength, for which the localization associated with NHSE is exactly canceled by the localization of in-gap states.
Interestingly, this transition can be regarded as a topological transition as well, because the spectral winding number with respect to the energy of the concerned edge state also jumps at this point.}  In particular,
the spectral winding number with respect to a reference energy $E_r$ is defined as follow \cite{gong2018,okuma2020,zhang2020a} :
\begin{equation}
\label{eqn:winding}
	\nu(E_r) = \frac{1}{2\pi}\oint_{BZ}\frac{d}{dk}\arg\det[H(k)-E_r]dk.
\end{equation}
This spectral winding number has been used to characterize the presence of the NHSE from the topological feature of the energetics on the complex plane.
{If there exists a reference energy $E_r$ such that the winding number with respect to $E_b$ is nonzero, then the NHSE presents in the system under open boundary condition and causes localization of the eigenstates at two different ends according to the sign of the winding number, i.e. positive sign indicates the localization to the left and negative sign indicates localization to the right.}
By taking the reference energy to be the energy of the edge state of which the localization direction is opposite to that of the NHSE, the spectral winding number, by definition,  changes where the PBC spectrum passes through the edge state, which is also the critical point for the delocalization of the same edge state.
{Furthermore, the non-zero spectral winding number with respect to the energy of the edge state also represents the regime where NHSE strength has overpowered the topological localization.
Therefore, the spectral winding number has a richer physics meaning now as it can be used to indicate the competition between the topological localization and the non-Hermitian pumping.}


\section{ Conclusion and discussion}

In this work, we have studied the localization properties of non-Hermitian topological edge states in three representative 1D models, including the non-Hermitian SSH model, non-Hermitian Floquet SSH model, and three-band extended SSH model.
At a critical NHSE strength,  the delocalization of topological edge states is observed in all three models.
In the first two models with chiral symmetry, it is described by that the energy (quasienergy) of the edge state is located at the EP of the PBC spectrum.
Numerical simulations are also presented in the Floquet SSH model, showing that \GJ{delocalized topological edges should be within the reach of experimental detection.
Furthermore, as shown in the third model with only conjugated-inversion symmetry, the delocalization of topological edge states does not occur at the EP, but only requires that
the real energy of the topological edge falls exactly on the PBC spectral loop.}
Such a requirement turns out to be the general condition for edge-state delocalization in all the models studied in this work.
Notably, the spectral winding number with respect to the energy of the edge state can serve as a tool to \GJ{reflect the inherent competition between the topological localization and the NHSE.  With this understanding the critical NHSE strength at which topological edge states are delocalized becomes the transition point at which the spectral winding number makes a jump.  }

\GJ{Delocalized topological edge states are fascinating because of the following two simultaneous features: they are gapped from the bulk and yet they are extended states.   Because such edge states are real valued and remain isolated from the bulk spectrum, it should be experimentally convenient to adiabatically manipulate them, without much nonadiabatic contamination from the bulk.  For example,  we envision that with the adiabatic following of the edge states, an  edge state initially localized at the left end can be adiabatically switched to an extended state, and then further to a localized state at the right end.  This kind of scenario may be highly useful for realizing edge state pumping or state transfer.   Since via this process an extended state is also adiabatically generated from a localized state, one may synthesize a giant Schr\"{o}dinger cat state starting from a localized edge state, if the non-Hermitian lattice indeed accommodates a quantum particle.  We plan to explore these interesting possibilities in our future work.}

\begin{acknowledgements}
J.G. acknowledges funding support by the Singapore Ministry of Education Academic Research Fund Tier-3 (Grant No. MOE2017-T3-1-001 and WBS No. R-144-000-425-592) and by the Singapore NRF Grant No. NRF-NRFI2017-04 (WBS No. R-144-000-378- 281).

\end{acknowledgements}

\bibliography{references}

\end{document}